\documentclass[12pt, prd, aps, showpacs, superscriptaddress, preprint]{revtex4-1}

\usepackage{amsmath}
\usepackage{mathtools}
\usepackage{graphicx}
\usepackage{mathrsfs}
\usepackage{hyperref}
\usepackage{subfigure}
\usepackage{slashed}

\begin{document}

\newcommand\OUT{\mathrm{out}}
\newcommand\IN{\mathrm{in}}
\newcommand\FV{\mathrm{FV}}
\newcommand\V{V}

\title{Effects of finite volume on the $K_L$-$K_S$ mass difference}

\newcommand\bnl{Brookhaven National Laboratory, Upton, NY 11973, USA}
\newcommand\cu{Physics Department, Columbia University, New York,
      NY 10027, USA}
\newcommand\riken{RIKEN-BNL Research Center, Brookhaven National
      Laboratory, Upton, NY 11973, USA}
\newcommand\soton{School of Physics and Astronomy, University of
  Southampton,  Southampton SO17 1BJ, UK}
\newcommand\sissa{SISSA, I-34136 Trieste and INFN Sezione di Roma La Sapienza, 00185 Roma, Italy}

\author{N.H.~Christ}\affiliation{\cu}
\author{X.~Feng}\affiliation{\cu}
\author{G.~Martinelli}\affiliation{\sissa}
\author{C.T.~Sachrajda}\affiliation{\soton}

\date{April 4, 2015}


\begin{abstract}
Phenomena that involve two or more on-shell particles are particularly sensitive to the effects of finite volume and require special treatment when computed using lattice QCD.  In this paper we generalize the results of L\"uscher, and Lellouch and  L\"uscher, which determine the leading order effects of finite volume on the two-particle spectrum and two-particle decay amplitudes to determine the finite-volume effects in the second order mixing of the $K^0$ and $\overline{K^0}$ states.  We extend the methods of Kim, Sachrajda and Sharpe to provide a direct, uniform treatment of these three, related, finite-volume corrections.  In particular, the leading, finite-volume corrections to the $K_L-K_S$ mass difference $\Delta M_K$ and the CP violating parameter $\epsilon_K$ are determined, including the potentially large effects which can arise from the near degeneracy of the kaon mass and the energy of a finite-volume, two-pion state.
\end{abstract}

\maketitle

\section{Introduction}

The mass difference $\Delta M_K$ between the  $K_L$ and $K_S$ mesons arises in the standard model at fourth order in the electro-weak coupling.  Its resulting small size ($3.484(6) \times 10^{-12}$\,MeV~\cite{Agashe:2014kda}) makes this quantity highly sensitive to new phenomena that lie outside the standard model.  The quantity  $\Delta M_K$ is the real part of the $K^0$-$\overline{K^0}$ mixing matrix element $M_{\overline{0}0}$.  The imaginary part of $M_{\overline{0}0}$ enters the parameter $\epsilon_K$ which describes indirect CP violation in the kaon system ($\epsilon_K=2.228 (11) \times 10^{-3}$~\cite{Agashe:2014kda}).  While both $\Delta M_K$ and $\epsilon_K$ are known precisely from experiment, their accurate calculation within the standard model poses an important challenge for the non-perturbative methods of lattice QCD.  

The quantity $\epsilon_K$ is dominated by short distance effects coming from the scale of the masses the $W$ boson and the top quark.  It can be computed in the standard model to an accuracy of approximately 5\% using QCD/electro-weak perturbation theory provided the $K^0$-$\overline{K^0}$ matrix element of a single $\Delta S=2$ four-quark operator  $\bigl(\overline{s}\gamma_\mu(1-\gamma^5)d\bigr)\bigl(\overline{s}\gamma_\mu(1-\gamma^5)d\bigr)$ has been computed using lattice methods.  Here $d$ and $s$ are the fields of the down and strange quarks respectively.  However, comparing the predictions of the standard model with the measured value of $\epsilon_K$ to greater accuracy will require the treatment of long-distance phenomena, at the energy scale of the charm quark mass and below.   In contrast, $\Delta M_K$ receives its largest contributions from phenomena whose energy scale lies at and below the charm quark mass, energies at which QCD/electro-weak perturbation theory cannot be used reliably~\cite{Brod:2011ty}.

Recently developed methods~\cite{Christ:2010zz, Yu:2011gk, Christ:2012np, Yu:2012nx, Christ:2012se, Yu:2013qfa, Bai:2014cva} promise to allow the calculation of these long-distance effects directly using lattice QCD.  This should permit percent-level tests of the standard model theory of CP violation and an increase of a factor of ten in the sensitivity of comparisons between the predictions of the standard model and experiment for $\Delta M_K$.  Since both of these calculations involve possible on-shell, intermediate two-pion states, they are susceptible to potentially significant finite-volume corrections.  It is the first-principles determination of these finite-volume effects which is the central topic of this paper.

The masses and matrix elements of single-hadron states computed in lattice QCD
are affected by finite-volume effects which decrease exponentially as $L$, the linear size
of the lattice, grows. Since the pion is the lightest hadron, in most cases the dominant
finite-volume corrections are proportional to $e^{-m_\pi L}$~\cite{Luscher:1986pf}.  However, if the energy of interest is above a two-particle threshold so that the two
particles can propagate without exponential suppression throughout the spatial volume, then power-law finite-volume corrections will result.  For the case of a single two-particle channel with energy below the threshold for three or more particles~\cite{Luscher:1990ux} or with two or more coupled two-particle channels~\cite{Hansen:2012tf}, it is possible to relate the finite-volume shift in the allowed two-particle energies to the infinite-volume scattering matrix.  This relation has proven to be a valuable tool, allowing the determination of scattering phase shifts from finite-volume energies which can be computed in lattice QCD.  

Similarly a two-particle decay matrix element can be computed using lattice QCD by exploiting this finite-volume quantization of the two-particle energies to adjust the energy of the two-particle final state to equal the mass of the decaying particle.  However, when relating the resulting on-shell, finite-volume matrix element to that in infinite volume the usual conversion factor, appropriate for non-interacting pions, requires an additional $O(1/L^3)$, finite-volume correction which can also be computed from the two-particle phase shifts~\cite{Lellouch:2000pv,Lin:2001ek}.  While originally presented for the case of the decay of a particle at rest, these results have been extended to the decay of a moving particle~\cite{Rummukainen:1995vs, Christ:2005gi, Kim:2005gf} and to the case of multi-channel, two-particle final states~\cite{Hansen:2012tf}.

Second-order weak amplitudes such as $\Delta M_K$ or $\epsilon_K$ represent a third topic in which on-shell, two-pion states can result in potentially significant finite-volume effects, similar to those analyzed by L\"uscher and by Lellouch and L\"uscher.  Recall that the mixing of the $K^0$ and $\overline{K^0}$ states is a text-book~\cite{Donoghue:1992dd} application of time-dependent perturbation theory and results in the formula for the complex mixing matrix $M_{ab}$
\begin{equation}
M_{ab} = {\cal P} \int_{2m_\pi}^\infty dE\sum_\alpha \frac{\langle K^a|H_W|\alpha,E\rangle \langle \alpha,E|H_W| K^b \rangle}{M_K-E},
\label{eq:mixing_infty}
\end{equation}
where the sum over the intermediate states $|\alpha, E\rangle$ includes an integral over the intermediate-state energy $E$ and a generalized sum over the other degrees of freedom represented by the label $\alpha$.  Here ${\cal P}$ indicates that the singularity from the denominator when $E=M_K$ is to be defined using the principal part prescription.  The indexes $a$ and $b$ take the values $0$ and $\overline{0}$ corresponding to the states $K^0$ and $\overline{K^0}$, respectively.   Since the center-of-mass coordinates are always included in this paper, we use the unintegrated, weak Hamiltonian density ${\cal H}_W$ in the right-most factor in Eq.~\eqref{eq:mixing_infty} to avoid introducing a delta function for three-momentum conservation.

As is shown in Ref.~\cite{Christ:2012se}, a finite-volume version of the right-hand side of this equation can be obtained from a Euclidean-space, lattice calculation:
\begin{equation}
M_{ab}^\V = \sum_n \frac{\langle K^a|H_W|n\rangle \langle n|H_W| K^b \rangle}{M_K-E_n},
\label{eq:mixing_FV}
\end{equation}
where now all the intermediate states $|n\rangle$ have discrete energies and the singularity in the denominator must be avoided by either choosing $L$ to prohibit intermediate states with energies degenerate with the kaon mass $M_K$ or by explicitly removing the singular term from the sum.

The first results~\cite{Christ:2010zz, Christ:2012np} for the finite-volume correction connecting the expressions in Eqs.~\eqref{eq:mixing_infty} and \eqref{eq:mixing_FV} were obtained using a generalization of the indirect method of Lellouch and L\"uscher and applied only to the special case that the energy of one of the two-pion intermediate states was tuned to match the mass of the kaon and that state was removed from the sum in Eq.~\eqref{eq:mixing_FV}.  A more general result, valid for an arbitrary sequence of finite-volume two-pion energies, was presented in Ref.~\cite{Christ:2014qaa}.  In this paper we will provide a derivation of this more general result which uses the techniques of Kim, Sachrajda and Sharpe (KSS)~\cite{Kim:2005gf}.  As we will see, this approach allows these three topics (finite-volume energy quantization, two-particle decay and second-order weak particle mixing) to be treated in a uniform way and provides a new, direct derivation of the results for the final two topics.  

Common to each of these three processes is the s-channel, two-particle-irreducible, $\pi-\pi$ scattering kernel which contains no finite-volume, power-law corrections.  In each case the same structure of finite-volume singularities in the center-of-mass energy is determined by this kernel, independent of whether the initial and final states are simply two pions or more complex products of a weak Hamiltonian acting on a kaon state.

We now briefly outline this approach.  We will study two closely related, Minkowski-space Green's functions $C_{\pi\pi}(E)$ and $C_K(E)^{ab}$ defined by
\begin{eqnarray}
C_{\pi\pi}(E) &=& \int d^3 r  \int_{-\infty}^\infty dt  
\langle 0|T\left\{\sigma_{\pi\pi}^\dagger(\vec 0, 0)\sigma_{\pi\pi}(\vec r, t)\right\}|0\rangle
 e^{i\left(\vec P\cdot \vec r - E t\right)}\label{eq:master_pipi} \\
C_K(E)^{ab} &=&-\frac{i}{2} \int d^4 x  
\langle 0|T\left\{{K^a}^\dagger(0)
\int dt_2 H_W(t_2)e^{iE t_2}  \int dt_1 H_W(t_1)e^{-iE t_1} 
K^b(x)\right\}|0\rangle
\nonumber \\
&&\hskip 1.0 in
e^{-i P_K\cdot x}\left. 
      \left(\frac{P_K^2 -M_K^2}{i}\right)^2\right|_{P_K^2=M_K^2}.
\label{eq:master_DM}
\end{eqnarray}
Here $\sigma_{\pi\pi}$ is a local, interpolating operator which can create a two-pion 
state from the vacuum.  The right-most factor in Eq.~\eqref{eq:master_DM}, $(P_K^2-M_K^2)^2$ amputates the two external kaon propagators ensuring that the right-hand side of that equation, evaluated on-shell at $P_K^2=M_K^2$, becomes a matrix element between physical initial and final kaons.  The correlation function $C_{\pi\pi}(E)$ contains an intermediate two-pion state carrying three momentum $\vec P$ and energy $E$, while the two-pion state which can appear in  $C_K(E)^{ab}$ carries momentum $\vec P_K$ and energy $E_K \pm E$ where the kaon four-momentum $P_K$ is given by $P_K = (\vec P_K, E_K)$ and the plus/minus sign results if the $H_W(x)$ vertex is contracted with the strange quark in the incoming/outgoing kaon.  For simplicity we specialize to the case of zero total momentum: $\vec P = \vec P_K = \vec 0$.  The indices $a$ and $b$ can take the values $0$ and $\overline{0}$  and for the second case, $K^{\overline{0}} \equiv \overline{K^0}$ .  Note that in addition to Eqs.~\eqref{eq:master_pipi} and \eqref{eq:master_DM}, all of the other equations and discussions in this paper are presented in Minkowski space.

By specifying the total incoming three-momentum with a spatial Fourier transform which introduces no explicit factors of the spatial volume and by imposing no additional constraint on the final three-momentum, we have defined both $C_{\pi\pi}(E)$ and $C_K(E)$ in such a way that they will have a well-defined infinite-volume limit.  It is the determination of the differences between each of these Green's functions when evaluated in finite and infinite volume:
\begin{equation}
C_X^\FV(E) = C_X^\V(E) - C_X^\infty(E) 
 \quad \mbox{for} \quad X=\pi\pi\;\;\mbox{or}\;\;K
\label{eq:corr_diff}
\end{equation}
which is the subject of this paper.

Kim, Sachrajda and Sharpe study $C_{\pi\pi}(E)$ and derive an explicit formula for $C_{\pi\pi}^\FV(E)$.  Because of the close relation between the results presented here and those in the earlier paper~\cite{Kim:2005gf} of KSS we will adopt the notation used in that paper.  In particular the superscript $\FV$ indicates the difference between a finite and an infinite volume result while the superscript $\V$ identifies the result that would be obtained in a finite volume $V$.  Up to exponentially small corrections, the difference $C_{\pi\pi}^\FV(E)$ comes entirely from on-shell, two-pion intermediate states and appears in essentially the same form in both $C_{\pi\pi}^\FV(E)$ and $C_K^\FV(E)$.  As originally recognized by L\"uscher, this difference can be written entirely in terms of infinite-volume quantities.  

In the approach of KSS, the poles of $C_{\pi\pi}^\FV(E)$ are directly related to the infinite-volume $\pi-\pi$ scattering phase shifts $\delta_l(E)$. However, these poles must be present only in the finite-volume Green's function $C_{\pi\pi}^\V(E)$ where they are located at the energies of the finite-volume two-pion states.  Thus, the infinite-volume $\pi-\pi$ phase shifts are constrained by the finite-volume $\pi-\pi$ energies.  For the case of the second Green's function $C_K(E)$ the same set of poles, now in the variable $E+M_K$, appear in $C_K^\FV(E)^{ab}$ and $C_K^\V(E)^{ab}$.  Equating their residues for the case $a=b=0$ gives directly the finite-volume corrections to the $K\to\pi\pi$ amplitude originally derived by Lellouch and L\"uscher.  Finally, when evaluated at $E=0$,  $C_K^\FV(E)^{\overline{0}0}$ reproduces the general formula for the finite-volume corrections to the off-diagonal, $K^0 - \overline{K^0}$ mixing matrix element $M_{\overline{0}0}$ which determines $\Delta M_K$ and $\epsilon_K$, given in Ref.~\cite{Christ:2014qaa}.

The remainder of this paper is organized as follows.  In Section~\ref{sec:FV_corr} we will present the expression obtained by KSS for the finite-volume correction $C_{\pi\pi}^\FV(E)$, summarize its derivation and recall how it can be used to obtain L\"uscher's relation between finite-volume, $\pi$-$\pi$ energies and infinite-volume $\pi$-$\pi$ scattering phase shifts.   Section~\ref{sec:LL} demonstrates that if a similar analysis is applied to the quantity $C_K^\FV(E)^{00}$ we can obtain directly the Lellouch-L\"uscher relation between finite- and infinite-volume $K\to\pi\pi$ decay amplitudes. In Section~\ref{sec:Delta_M} we observe that when evaluated at $E=0$, $C_K^\FV(E)^{ab}$ gives the difference $M^\FV_{ab} = M_{ab}^\V - M_{ab}^\infty$ that is needed to remove finite-volume effects from the quantity $M_{ab}^\V$ --- a quantity which can be calculated using lattice methods. Finally, Section~\ref{sec:conclusion} contains some concluding remarks.

\section{Finite-volume corrections to \boldmath{$C_{\pi\pi}(E)$} and finite-volume energy quantization}
\label{sec:FV_corr}

\begin{figure}[!htp]
\centering
\includegraphics[width=0.9\textwidth]{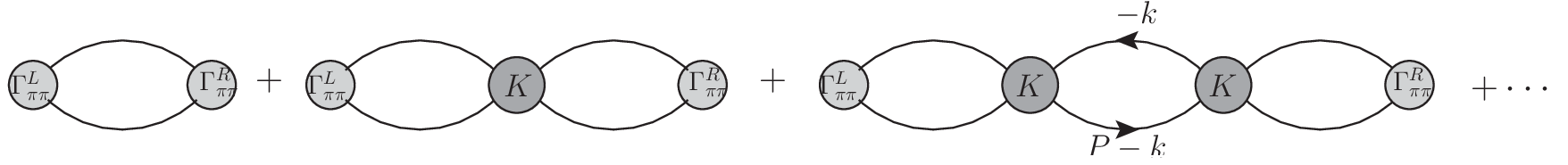}
\caption{A graphical representation of the decomposition of the amplitude $C_{\pi\pi}(E)$ into a series of terms, each with a specific number of maximal, two-particle irreducible subgraphs.  The shaded kernel $K$ with four external pion lines represents the amputated, two-particle irreducible $\pi$-$\pi$ scattering amplitude while the right and left vertices labeled $\Gamma_{\pi\pi}^L$ and $\Gamma_{\pi\pi}^R$ correspond to the amputated, two-particle irreducible portions of each graph which contain the interpolating operator $\sigma_{\pi\pi}$ acting on the vacuum.  Free particle propagators are used to join the subgraphs shown, with the appropriate factors $Z_\pi(k^2)$ included in the $\Gamma_{\pi\pi}^L$ and $K$ amplitudes as described in the text.}
\label{fig:2PI}
\end{figure}

In this section we review the approach of Kim, Sachrajda and Sharpe to determine the finite-volume corrections to the correlation function $C_{\pi\pi}(E)$ given in Eq.~\eqref{eq:master_pipi}.  Their starting point is the usual diagrammatic expansion of $C_{\pi\pi}(E)$ into products of amputated, s-channel, two-particle irreducible, four-particle correlation functions $K$ connected by pairs of single-particle propagators as shown in Figure~\ref{fig:2PI}.  Here s-channel, two-particle irreducible means that the kernel $K$ is constructed from graphs which cannot be divided into two disconnected components, one containing the two input pion lines and the other the output pion lines by cutting only two intermediate pion lines.  This graphical sum can be expressed algebraically by the following equation in which the resulting geometric series is evaluated:
\begin{eqnarray}
C'_{\pi\pi}(E) &=&
    \Gamma_{\pi\pi}^L S_2 \Gamma_{\pi\pi}^R
  +\Gamma_{\pi\pi}^L S_2 K S_2 \Gamma_{\pi\pi}^R
  + \ldots \\
   &=&\Gamma_{\pi\pi}^L S_2\frac{1}{1-K S_2}\Gamma_{\pi\pi}^R.
\label{eq:LS_equation}
\end{eqnarray}
The prime indicates that $C'_{\pi\pi}(E)$ includes only terms with one or more two-pion intermediate states.  The final and initial factors $\Gamma_{\pi\pi}^L$ and $\Gamma_{\pi\pi}^R$ correspond to amputated, two-particle irreducible amplitudes containing the two-pion interpolating operators $\sigma_{\pi\pi}^\dagger$ and $\sigma_{\pi\pi}$ while $S_2$ represents the product of the two free scalar propagators for the two intermediate particles and $K$ the two-particle irreducible, $\pi$-$\pi$ scattering kernel.  Each matrix product in Eq.~\eqref{eq:LS_equation} in principle involves integrations over the four-momenta of two pions.  However, we reduce this eight-dimensional integration volume to the four-dimensional volume of a single pion four-momentum by using the conservation of the total energy and momentum.  As mentioned earlier, we work in the $\pi$-$\pi$ rest frame with $\vec P = 0$.

The usual graphical decomposition of the amplitude $C_{\pi\pi}(E)$ suggested by Figure~\ref{fig:2PI} involves a full scalar propagator, $S(k^2)$, corresponding to each meson line in that figure.   In obtaining Figure~\ref{fig:2PI} and Eq.~\eqref{eq:LS_equation} we removed a factor $Z_\pi(k^2)$, which is an analytic function of $k^2$ for $k^2 < (3 m_\pi)^2$, from the full propagator $S(k^2)$, 
\begin{equation}
S(k^2) = \frac{iZ_\pi(k^2)}{k_0^2 - \vec k\,^2 - m_\pi^2 + i\epsilon}
\end{equation}
and absorbed those two factors of $Z_\pi$ into the definition of the two-particle irreducible scattering kernel $K$ and the left-hand vertex $\Gamma_{\pi\pi}^L$.  So that free scalar propagators are represented by the meson lines in Figure~\ref{fig:2PI} and appear in the product $S_2$ in Eq.~\eqref{eq:LS_equation}.  (For simplicity, we will normalize the scalar fields $\phi$ so that $Z_\pi(m_\pi^2)=1$.)

We should point out, that while we are interested in establishing relations between physical quantities in QCD, it is not known how to make this sort of diagrammatic expansion in QCD where the relevant graphs are composed of quark and gluon propagators, not the pion propagators which are used in the classification upon which Figure~\ref{fig:2PI} is based.  Consequently, the present and earlier studies relating finite-volume effects to physical, infinite-volume scattering properties are derived for an artificial theory in which the pions are elementary particles, not quark-anti-quark bound states.  We then assume that general relations which are found in such an artificial theory are universal since we are describing long-distance effects and must also be obeyed by corresponding quantities in QCD.

All power-law, finite-volume corrections present in $C_{\pi\pi}^\V(E)$ come from the sums over the discrete, two-particle, spatial momenta associated with each factor of $S_2$ in Eq.~\eqref{eq:LS_equation}.  A typical such sum can be written
\begin{eqnarray}
{\cal F} &=& \int_{-\infty}^\infty dk_0 \sum_{\vec k} X_L(\vec k,k_0)
        \frac{1}{\bigl(k_0^2 -\vec k^2 -m_\pi^2+i\epsilon\bigr)} \label{eq:int_k_0} \\ 
&& \hskip 1.0 in\cdot \frac{1}{\bigl((E-k_0)^2 -\vec k^2 -m_\pi^2+i\epsilon\bigr)}
X_R(\vec k,k_0),
\nonumber
\end{eqnarray}
where for periodic boundary conditions in a cubic box of side $L$, $\vec k = 2\pi(n_1,n_2,n_3)/L$ where the ${n_i}$, ${1 \le i \le 3}$ are integers.   Here $X_L(\vec k,k_0)$ and $X_R(\vec k,k_0)$ represent the left- and right-hand factors in the product in which this particular discrete sum appears.  Following L\"uscher~\cite{Luscher:1986pf} and KSS, we rearrange the $k_0$ integration contour in Eq.~\eqref{eq:int_k_0} to isolate that part of the right hand side of Eq.~\eqref{eq:int_k_0} which is not analytic in the summation variable $\vec k$.  

For example, we can begin with a value of $E$ between 0 and $m_\pi$.  In that case the Minkowski-space integral over $k_0$ can be Wick rotated counter-clockwise to a contour along the imaginary axis which lies to the left of explicit $k_0$ poles at $\omega_k$ and $E+\omega_k$ but to the right of poles at $-\omega_k$ and $-\omega_k+E$, where $\omega_k = \sqrt{k^2 + m_\pi^2}$.  This contour can then be shifted to the right of the pole at $k_0 = \omega_k$ and Cauchy's theorem used to express the original integral as the sum of the shifted integral and an extra term evaluated at the $k_0=\omega_k$ pole.  We can then increase $E$ into the region of interest, $2m_\pi < E < 4m_\pi$.  In this region it is only this $k_0=\omega_k$ pole contribution which is non-analytic in the three-momentum $\vec k$.  This singularity corresponds to the $1/(k_0+\omega_k-E)$ pole from the second propagator evaluated at $ k_0=\omega_k$.  The power-law, finite-volume corrections can be then obtained by applying the Poisson summation formula to this singularity in the product of two free-particle propagators which occurs when both particles are on-shell.

Using the Poisson summation formula KSS obtain an expression for the power-law, finite-volume correction to the amplitude $\cal F$ which involves only infinite-volume, on-shell quantities:
\begin{eqnarray}
{\cal F}^\FV = \sum_{l_2,m_2}\sum_{l_1,m_1} X_L(q)_{l_2,m_2}F(q)_{l_2m_2,l_1,m_1}X_R(q)_{l_1,m_1},
\label{eq:KSS_FV-1}
\end{eqnarray}
where $q=\sqrt{(E/2)^2-m_\pi^2}$ is the energy-conserving momentum of each of the on-shell, intermediate pions.  The factors $X_{L/R}(q)_{l,m}$ are the amplitudes $X_{L/R}(\vec k,k_0)$ evaluated on-shell and projected onto definite angular momentum eigenstates.  For example,
\begin{equation}
X_R(q)_{l_1,m_1} = \int d\Omega\, Y_{l_1 m_1}^*(\theta,\phi) X_R\bigl(\hat \Omega(\theta,\phi) q,\omega_q\bigr),
\end{equation}
where $\hat \Omega(\theta,\phi)$ is a unit vector whose direction is specified by the usual polar coordinates $\theta, \phi$ while the single-pion energy $\omega_q = \sqrt{q^2+m_\pi^2}$.  The finite-volume correction matrix $F_{l_2m_2,l_1,m_1}$ is given by:
\begin{eqnarray}
F(q)_{l_2 m_2,l_1,m_1} =  \frac{q\sqrt{4\pi}}{16\pi\omega_q} 
          \sum_{l,m}\left\{ \delta_{l0}\delta_{m0}  
                         -i\frac{4\pi}{q^{l+1}} c_{lm}(q^2)\right\}
\label{eq:KSS_FV-2} 
\left\{\int d \Omega_{\hat p} Y_{l_2,m_2}  Y_{l,m}  Y^*_{l_1,m_1}\right\},
\end{eqnarray}
where
\begin{equation}
c_{lm}(q^2) = \frac{1}{L^3} \sum_{\vec k} 
                       \frac{e^{\alpha(q^2 -  k^2)}}{q^2-k^2} k^l \sqrt{4\pi}
                                  Y_{lm}(\theta_{\hat k},\phi_{\hat k})
  -\delta_{l0} {\cal P}\int \frac{d^3k}{(2\pi)^3}\frac{e^{\alpha(q^2-k^2)}}{q^2-k^2}
\label{eq:KSS_FV-3}
\end{equation}
and $k=|\vec k|$.  The exponential factor was introduced by KSS to make the sum over $\vec k$ convergent and $c_{lm}(q^2)$ should be evaluated in the limit $\alpha\to0^+$.  (The Eqs.~\eqref{eq:KSS_FV-1}-\eqref{eq:KSS_FV-3} above are equivalent to Eqs. (21)-(23) and (42) in KSS.)

\begin{figure}[!htp]
\centering
\includegraphics[width=0.9\textwidth]{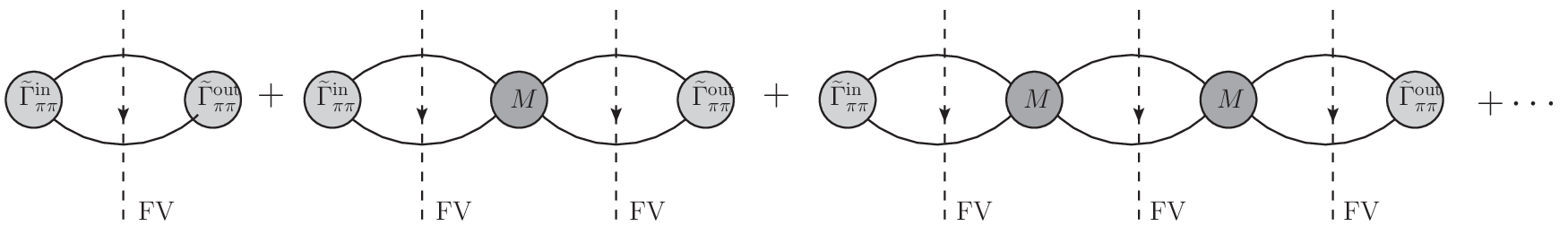}
\caption{A graphical representation of the decomposition of the finite-volume correction to the amplitude $C_{\pi\pi}(E)$ into a series of terms, each with a specific number of finite-volume corrections to the sum over two-particle states.  The shaded kernel $M$ is the infinite-volume $\pi$-$\pi$ scattering amplitude while the right and left vertices labeled $\widetilde{\Gamma}_{\pi\pi}^\IN$ and $\widetilde{\Gamma}_{\pi\pi}^\OUT$ correspond to the complete amputated, infinite-volume Green's functions containing two external pion lines and the interpolating operators $\sigma_{\pi\pi}^\dagger$ or $\sigma_{\pi\pi}$ acting on the vacuum.  These correspond to the column and row vectors $\widetilde{\Gamma}_{\pi\pi}^\OUT$ $\widetilde{\Gamma}_{\pi\pi}^\IN$ defined in Eqs.~\eqref{eq:out_state} and \eqref{eq:in_state}.}
\label{fig:2PI_FV}
\end{figure}

If each of the sums over the two-particle momenta represented in Figure~\ref{fig:2PI} is written as the sum of the infinite-volume result and the finite-volume correction given in Eqs.~\eqref{eq:KSS_FV-1}-\eqref{eq:KSS_FV-3}, then the graphical sum shown in Figure~\ref{fig:2PI} can be rearranged and the sum over the infinite-volume terms performed first.  This sum over a series of infinite-volume, two-particle contributions will result in the infinite-volume $\pi$-$\pi$ scattering amplitude $M$.  The remaining sum over various numbers of finite-volume corrections can then be organized as a second geometric series as shown in Figure~\ref{fig:2PI_FV} and represented algebraically as:
\begin{eqnarray}
C_{\pi\pi}^\FV(E) &=&
       \widetilde{\Gamma}_{\pi\pi}^\IN\Bigl(-\frac{1}{2}F\Bigr) \widetilde{\Gamma}_{\pi\pi}^\OUT
    + \widetilde{\Gamma}_{\pi\pi}^\IN\Bigl(-\frac{1}{2}F\Bigr)iM\Big(-\frac{1}{2}F\Big)
                                                                \widetilde{\Gamma}_{\pi\pi}^\OUT + \ldots \\
&=&-\frac{1}{2}\widetilde{\Gamma}_{\pi\pi}^\IN  F\frac{1}{1+\frac{i}{2} MF}
\widetilde{\Gamma}_{\pi\pi}^\OUT.
\label{eq:FV_series_pipi}
\end{eqnarray}
Here $F$ is the matrix defined in Eq.~\eqref{eq:KSS_FV-2}, $M$ is the infinite-volume, two-pion scattering matrix while $\widetilde{\Gamma}_{\pi\pi}^{\OUT/\IN}$ are column/row vectors describing the infinite-volume, on-shell coupling of two pions to the operator $\sigma_{\pi\pi}$.   In contrast to the two-particle irreducible vectors $\Gamma_{\pi\pi}^L$ and $\Gamma_{\pi\pi}^R$ which appear in Eq.~\eqref{eq:LS_equation}, the vectors $\widetilde{\Gamma}_{\pi\pi}^\IN$ and $\widetilde{\Gamma}_{\pi\pi}^\OUT$ are two-particle reducible and contain the full $\pi-\pi$ interaction present in the final or initial $\pi-\pi$ state.

These matrices and vectors with $l$ and $m$ angular momentum indices can be written in terms of standard infinite-volume Green's functions as follows.  The $\pi-\pi$ scattering amplitude $M_{lm}$ can be obtained from the amputated, two-particle scattering amplitude
\begin{eqnarray}
 && \prod_{i=1}^4\left\{ \frac{p_i^2-m^2}{i}\right\}\prod_{i=1}^4 \left\{\int d^4 x_i \right\}
  e^{i(p_4 x_4+p_3-p_2 x_2 -p_1 x_1)} 
   \bigl\langle \phi(x_4)\phi(x_3)\phi(x_2)\phi(x_1)\bigr\rangle
\nonumber \\
&& \hskip 1.5 in = (2\pi)^4\delta^4(p_4+p_3-p_2-p_1) M(p_4,p_3,p_2,p_1)
\end{eqnarray}
by putting the initial and final particles on shell and projecting onto angular momentum eigenstates:
\begin{eqnarray}
i M(q)_{lm}\delta_{l'l}\delta_{m'm} 
    &\equiv& \frac{1}{4\pi}\int Y_{l'm'}^*(\hat q\,') d\Omega_{\hat q\,'} \int Y_{lm}(\hat q) d\Omega_{\hat q} \label{eq:M_vec2ang} \\
    &&\hskip 0.2 in M\bigl((\vec q\,',\omega_q),
                      (-\vec q\,',\omega_q),(\vec q,\omega_q),(-\vec q,\omega_q)\bigr),
\nonumber
\end{eqnarray}
where the vectors $\vec q = q\hat q$ and $\vec q\,'=q\hat q'$ are each proportional to the unit vectors over whose directions we are integrating.  With this KSS choice of normalization, $M(q)_{lm}$ can be expressed in terms of the scattering phase shifts using the formula:
\begin{equation}
M(q)_{lm} = 32\pi\frac{\omega_q}{q}\frac{\bigl(e^{2i\delta_l}-1\bigr)}{2i},
\label{eq:M_lm-2}
\end{equation}

Finally the column/row vectors $\widetilde{\Gamma}_{\pi\pi}^{\OUT/\IN}$ are given by:
\begin{eqnarray}
\widetilde{\Gamma}_{\pi\pi}^\OUT(E)_{lm} &=& 
\frac{1}{\sqrt{4\pi}} \int d\Omega_{\hat q} Y^*(\hat q)_{lm} \prod_{i=1,2} \left\{\int d^4 x_i 
e^{ip_ix_i} \frac{p_i^2-m^2}{i}\right\}
                \Bigl\langle T\Bigl\{\phi(x_1)\phi(x_2) \sigma_{\pi\pi}(\vec 0, 0)\Bigr\}\Bigr\rangle 
\nonumber \\
&=& \frac{1}{\sqrt{2\pi}}2\omega_q \bigl\langle\pi\pi^\OUT(\vec 0,q,l,m)\bigl|\sigma_{\pi\pi}(\vec 0, 0)\bigr |0\bigr\rangle 
 \label{eq:out_state}
\end{eqnarray} 
and 
\begin{eqnarray}
\widetilde{\Gamma}_{\pi\pi}^\IN(E)_{lm} &=& 
\frac{1}{\sqrt{4\pi}} \int d\Omega_{\hat q} Y(\hat q)_{lm} \prod_{i=1,2} \left\{\int d^4 x_i e^{-ip_ix_i} \frac{p_i^2-m^2}{i}\right\}
                \Bigl\langle T\Bigl\{\sigma_{\pi\pi}(\vec 0, 0)\phi(x_1)\phi(x_2)\Bigr\}\Bigr\rangle 
\nonumber \\
&=& \frac{1}{\sqrt{2\pi}}2\omega_q \;\bigl\langle0\bigl|\sigma_{\pi\pi}(\vec 0, 0)\bigr |\pi\pi^\IN(\vec 0,q,l,m)\bigr\rangle, 
 \label{eq:in_state}
\end{eqnarray} 
where the momenta $p_1$ and $p_2$ are on shell and given by $p_{1/2} = (\pm q \hat q,\omega_q)$.  The angular momentum eigenstates $|\pi\pi^{\OUT/\IN}(\vec P,q,l,m)\rangle$ obey the normalization condition:
\begin{equation}
\bigl\langle\pi\pi^{\OUT/\IN}(\vec P',q',l',m')|\pi\pi^{\OUT/\IN}(\vec P,q,l,m) \bigr\rangle
  = \frac{(2\pi)^6}{q^2}\delta^3(\vec P' -\vec P)\delta(q'-q)\delta_{l'l}\delta_{m'm}.
\label{eq:pipi_norm}
\end{equation}

Following KSS, we use Eq.~\eqref{eq:FV_series_pipi} to obtain L\"uscher's finite-volume quantization condition~\cite{Luscher:1986pf, Luscher:1990ux} relating the energies $E_n$ of the discrete, finite-volume two-particle states to the infinite-volume $\pi-\pi$ scattering phase shifts which appear in Eq.~\eqref{eq:M_lm-2}.  Since each discrete, finite-volume energy $E_n$ must correspond to a pole in the finite-volume correlator $C_{\pi\pi}^\V(E)$ at $E=E_n$ and no such poles are present in the infinite volume correlator, the finite-volume correction $C_{\pi\pi}^\FV(E)$ must also contain these poles, which can be recognized as roots of the determinant of the matrix $1+\frac{i}{2}MF$ which appears in the denominator of Eq.~\eqref{eq:FV_series_pipi}.

For the case in which only the $s$-wave $\pi$-$\pi$ phase shift is non-zero the determinant of the matrix  $1+\frac{i}{2}MF$ is proportional to a known function of  $\delta_0(E)$: 
\begin{eqnarray}
\mathrm{Det}(1+\frac{i}{2}MF) &\propto& 1+\frac{i}{2}M_{00}F_{00,00} \\
   &=&\frac{\cot(\phi)+\cot(\delta_0)}{\cot(\delta_0) -i}
\label{eq:L_denom}
\end{eqnarray}
where the angle $\phi(E)$, originally introduced by L\"uscher, is defined in our context by
\begin{equation}
\cot(\phi) = \frac{4\pi}{q}c_{00}(E).
\end{equation} 
The finite-volume, energy eigenvalues $E_n$ must then be zeros of the function given Eq.~\eqref{eq:L_denom}:
\begin{equation}
\sin\bigl(\phi(E_n)+\delta_0(E_n)) = 0 
\end{equation}
or 
\begin{equation}
\phi(E_n) + \delta_0(E_n) = n\pi, 
\end{equation}
where $n$ is an integer.  For completeness, we also give L\"uscher's original expression for $\phi(E)$:
\begin{equation}
\tan(\phi(E)) = -\frac{\pi^{3/2}\bigl(\frac{qL}{2\pi}\bigr)} 
                                                          {{\cal Z}_{00}\bigl(1,\bigl(\frac{qL}{2\pi}\bigr)^2\bigr)}
\quad\mbox{where} \quad
{\cal Z}_{00}(s,x)
 = \frac{1}{\sqrt{4\pi}} \sum_{\vec n \in Z^3} \frac{1}{\bigl(\vec n^2 - x\bigr)^s}
\label{eq:phi_Luscher}
\end{equation}
In contrast to Eq.~\eqref{eq:KSS_FV-3} where an explicit exponential regulator has been added to make the sum over $\vec n$ finite, the function ${\cal Z}_{00}(s,x)$ specified in Eq.~\eqref{eq:phi_Luscher} is to be understood as defined for complex $s$ when Re$(s) > 3/2$ and then analytically continued to the point of interest, $s=1$.

In the next two sections we will go beyond the discussion presented in KSS and examine  how these finite-volume poles at $E=E_n$ enter the finite-volume corrections to the second correlator $C_K(E)$, given in Eq.~\eqref{eq:master_DM}.  Equating the residues of the poles in the finite-volume amplitude $C_K^\V(E)$ with those in the infinite-volume quantity $C_K^\FV(E)$  will  provide a direct derivation of the familiar Lellouch-L\"uscher relation between the finite- and infinite-volume $K\to\pi\pi$ decay matrix elements.  Using the equation directly will give the finite-volume correction for the second-order weak amplitude which describes $K^0 - \overline{K^0}$ mixing.

\section {Finite-volume corrections to \boldmath{$K\to\pi\pi$ decay}}
\label{sec:LL}

The  Lellouch-L\"uscher relation between the finite- and infinite-volume $K\to\pi\pi$ decay matrix elements follows easily from an application of the methods of the previous section to the $K^0-K^0$ correlation function $C_K(E)^{00}$ defined in Eq.~\eqref{eq:master_DM}.  As in Sec.~\ref{sec:FV_corr}, we study the difference $C_K^\FV$ defined in Eq.~\eqref{eq:corr_diff}.  Except for the column and row vectors, the finite-volume correction $C_K^\FV(E)^{00}$ can be expressed as a geometric series in the finite-volume correction matrix $F(q)$ that is identical to that given in Eq.~\eqref{eq:FV_series_pipi} for $C_{\pi\pi}^\FV$:
\begin{equation}
C_K^\FV(E)^{00} = -\frac{i}{2}\widetilde{\Gamma}_K^\IN  
                 F\frac{1}{1+\frac{i}{2}MF}\widetilde{\Gamma}_K^\OUT.
\label{eq:FV_series_K}
\end{equation}
Here the column and row vectors $\Gamma_K^\OUT$ and $\Gamma_K^\IN$ are given by analogues of Eqs.~\eqref{eq:out_state} and \eqref{eq:in_state}.  We can determine $\Gamma_K^\OUT$ as follows:
\begin{eqnarray}
\widetilde{\Gamma}_K^\OUT(E)_{lm}^{00} &=& 
\frac{1}{\sqrt{4\pi}} \int d\Omega_{\hat q} Y^*(\hat q)_{lm} 
\prod_{i=1,2} \left\{\int d^4 x_i e^{ip_ix_i} \frac{p_i^2-m^2}{i}\right\}
\int d^4 x e^{-ip_K\cdot x} \frac{p_K^2-M_K^2}{i}
\nonumber \\
&&\hskip 0.5 in  \Bigl\langle T\Bigl\{\phi(x_1)\phi(x_2) {\cal H}_W(\vec r=\vec 0,t=0) K^0(x)\Bigr\}\Bigr\rangle
\nonumber \\
&=& \frac{1}{\sqrt{2\pi}}2\omega_q
\bigl\langle\pi\pi^\OUT(\vec 0,q,l,m)\bigl|{\cal H}_W\bigr |K(\vec p_K=0)\bigr\rangle \sqrt{2M_K} 
 \label{eq:out_state_K}
\end{eqnarray} 
where, as in Eqs.~\eqref{eq:out_state} and \eqref{eq:in_state}, the momenta $p_1$ and $p_2$ are on shell and given by $p_{1/2} = (\pm q \hat q,\omega_q)$ and ${\cal H}_W$ is the Hamiltonian density whose spatial integral is the Hamiltonian $H_W$.
Because of the energy injected into the weak interaction Hamiltonian in Eq.~\eqref{eq:master_DM} and our choice of amplitude described below, $2\omega_q=M_K+E$.  A similar expression can be obtained for $\widetilde{\Gamma}_K^\IN(E)_{lm}$:
\begin{eqnarray}
\widetilde{\Gamma}_K^\IN(E)_{lm} &=& \frac{1}{\sqrt{2\pi}}2\omega_q 
\bigl\langle K(\vec p_K=0) \bigl|{\cal H}_W\bigr |\pi\pi^\IN(\vec 0,q,l,m)\bigr\rangle \sqrt{2M_K}, 
 \label{eq:in_state_K}
\end{eqnarray} 
where, as in Eq.~\eqref{eq:out_state_K}, $M_K+E=2\omega_q$.  By introducing the injection and extraction of the energy $E$ by $H_W(t)$ in Eq.~\eqref{eq:master_DM} we have created a distinction between these two otherwise identical operators: one, $H_W(t_1)$, injects the energy $E$ and the other, $H_W(t_2)$ extracts $E$.  As a result we can distinguish two classes of contractions: a first in which $H_W(t_1)$ is contracted with the strange quark present in the initial kaon state and a second in which that initial strange quark is contracted with $H_W(t_2)$.  To simplify the following discussion we will consider only the first class of contractions (the case where the intermediate $\pi\pi$ energy is $M_K+E$).  We will also remove the usual factor 1/2 shown in Eq.~\eqref{eq:master_DM} so that in the limit $E \to 0$ we will recover the correctly normalized, second-order matrix element from this single class of contractions. 

In obtaining Eqs.~\eqref{eq:out_state_K} and \eqref{eq:in_state_K} we have also adopted the usual treatment of the center-of-mass variables, designed to avoid introducing unnecessary differences between finite- and infinite-volume amplitudes.  For infinite-volume matrix elements we use the Hamiltonian density ${\cal H_W}$ and states that are normalized to $(2\pi)^3$ times a delta function in the center-of-mass momentum while for finite-volume matrix elements we use states normalized to unity and the integrated Hamiltonian $H_W=\int d^3 r {\cal H}_W(\vec r)$.

As in the derivation of the L\"uscher quantization condition, we can argue that the poles which occur in $C^\FV_K(E)$ must arise from the finite-volume amplitude $C^\V_K(E)$ and equating their residues in $C^\FV_K(E)$ and $C^\V_K(E)$ as required by Eq.~\eqref{eq:corr_diff} will then relate the finite- and infinite-volume $\langle \pi\pi|H_W|K^0\rangle$ matrix elements.  Thus, we should equate residues of the poles at  $E=E_n$ in the right- and left-hand sides of the following equation:
\begin{eqnarray}
\frac{\bigl|\prescript{\V}{}{\langle} \pi\pi|H_W|K^0\rangle\bigr|^2}{E+M_K-E_n} &=&
 \frac{\omega_q q }{16\pi^2}\frac{\langle K^0|{\cal H}_W|\pi\pi^\IN{\rangle}^\infty e^{i(\phi-\delta_0)}
                  \prescript{\infty}{}{\langle}\pi\pi^\OUT|{\cal H}_W| K^0\rangle}
{\sin(\phi+\delta_0)} \label{eq:LL_1} \\
 &=&
 \frac{\omega_q q }{16\pi^2}\frac{\bigl|\prescript{\infty}{}{\langle}\pi\pi^\OUT|{\cal H}_W| K^0\rangle\bigr|^2 e^{i(\phi+\delta_0)}}
{(E+M_K-E_n)\frac{d(\phi+\delta)}{dE} \cos{(\phi+\delta_0)}}
\label{eq:LL_2}
\end{eqnarray}
where the left-hand side of Eq.~\eqref{eq:LL_1} is the relevant term in the finite volume amplitude $C^\V_K(E)$ while the right-hand side of that equation is the KSS expression for the difference between the finite- and infinite-volume correlation functions, $C^\FV_K(E)$.  In going from Eq.~\eqref{eq:LL_1} to Eq.~\eqref{eq:LL_2} we have extracted the phase factors associated with the in and out states (often referred to as the Watson phase factors) from the two matrix elements in the numerator of the right-hand side of  Eq.~\eqref{eq:LL_1} and combined them with the expression obtained by expanding the denominator at the $E+M_K=E_n$ pole. This gives the usual Lellouch-L\"uscher relation between the finite-volume and infinite-volume decay matrix elements:
\begin{equation}
\bigl|\prescript{\infty}{}{\langle}\pi\pi^\OUT|{\cal H}_W(\vec 0)| K^0\rangle\bigr|^2 = \frac{16 \pi^2}{\omega_q q} \bigl|\langle \pi\pi|H_W|K^0\rangle^\V\bigl|^2\frac{d(\phi+\delta)}{dE}.
\label{eq:LL_3}
\end{equation}
The form of this equation depends on the normalization condition that we have adopted for the $s$-wave infinite-volume $\pi\pi$ scattering state, $|\pi\pi^\OUT\rangle^\infty$ given in Eq.~\eqref{eq:pipi_norm}.  This equation reduces to the original Lellouch-L\"uscher relation if our normalization conventions for this infinite-volume state are converted to theirs.

Recall that in our definition of $C_K(E)$, the energy $E$ is carried by the operator $H_W$ while the kaon is on-shell.  Thus, as observed in Ref.~\cite{Lin:2001ek}, Eq.~\eqref{eq:LL_3} holds not only in the case of an energy-conserving decay matrix element when the volume has been adjusted so that $E=E_n-M_K=0$ but for general values of $E+M_K$ that lie above the two-pion threshold but are sufficiently small that mixing with multi-pion states can be neglected.  This result can also be easily generalized to relate the finite- and infinite-volume matrix elements of a variety of operators between single-particle initial states and two-pion final states~\cite{Lin:2001ek}.

\section {Finite-volume corrections to \boldmath{$\Delta M_K$} and \boldmath{$\epsilon_K$}}
\label{sec:Delta_M}

Finally we determine the finite-volume corrections to the $\overline{K^0} - K^0$ matrix element $M_{\overline{0}0}$ given in Eq.~\eqref{eq:mixing_infty}.  The real part of $M_{\overline{0}0}$ determines $\Delta M_K$ and its imaginary part gives $\epsilon_K$.  We begin with Eq.~\eqref{eq:corr_diff} for the case $X=K$ and $ab=\overline{0}0$ written as:
\begin{equation}
C_K^\infty(E=0)^{\overline{0}0} = C_K^\V(E=0)^{\overline{0}0} - 
                                                            C_K^\FV(E=0)^{\overline{0}0}.
\label{eq:DM_1}
\end{equation}
Since the quantities $C_K(E=0)^{\overline{0}0}$ have been chosen so that their dispersive parts correspond precisely to $M_{\overline{0}0}$ for both the finite- and infinite-volume cases, we can simply substitute the KSS result for $C_K^\FV(E=0)^{\overline{0}0}$ into Eq.~\eqref{eq:DM_1} to obtain a formula which provides the desired relation between $M_{\overline{0}0}^\infty$ and $M_{\overline{0}0}^\V$:
\begin{eqnarray}
\int_{2m_\pi}^\infty dE \sum_\alpha 
\frac{\langle \overline{K^0}|H_W|\alpha,E{\rangle}^\infty 
         \prescript{\infty}{}{\langle} \alpha,E|{\cal H}_W(\vec 0)| K^0\rangle}
        {M_K-E+i\epsilon} 
&=& \sum_n \frac{\langle \overline{K^0}|H_W|n{\rangle}^\V 
         \prescript{\V}{}{\langle}n|H_W| K^0\rangle}
        {M_K-E_n} 
\label{eq:DM_2} \\ \nonumber
&& \hskip -2.3 in - \frac{\omega_q q}{16\pi^2}\Bigl( \cot\bigl(\phi(M_K)+\delta_0(M_K)\bigr) + i\Bigr)
 \\ \nonumber
 && \hskip -1.8 in       \cdot\bigl\langle \overline{K^0}\bigl|{\cal H}_W\bigr|(\pi\pi)^\OUT,E=M_K{\bigr\rangle}^\infty 
                   \prescript{\infty}{}{\bigl\langle}(\pi\pi)^\OUT,E=M_K\bigl|{\cal H}_W\bigr| K^0\bigr\rangle,
\end{eqnarray}
where in the correction term in the third line of this equation we have removed a factor of $e^{2i\delta_0}$ from the matrix element of what was an ``in'' state so that the product of bra and ket ``out'' states results and only the phase associated with the two potentially CP-violating matrix elements of ${\cal H}_W$ appears in the correction term.  The Watson phase has been written explicitly.

The denominator $M_K-E+i\epsilon$ of the left-hand side of Eq.~\eqref{eq:DM_2} gives two terms: the principal part, which defines $M_{\overline{0}0}$, as well as a $-i\pi\delta(E-M_K)$ term.  The delta function term is not present in the finite-volume portion  $M_{\overline{0}0}^\V$ of the right hand side and instead corresponds to the $+i$ term within the large curved brackets:
\begin{eqnarray}
-i\pi \sum_\alpha 
\langle \overline{K^0}|H_W|\alpha,M_K{\rangle}^\infty 
         \prescript{\infty}{}{\langle} \alpha,M_K|{\cal H}_W| K^0\rangle &=& \label{eq:check}
\\ \nonumber
&&\hskip -2.0 in  - i\frac{\omega_q q}{16 \pi^2}\bigl\langle \overline{K^0}\bigl|{\cal H}_W\bigr|(\pi\pi)^\OUT,M_K{\bigr\rangle}^\infty 
                   \prescript{\infty}{}{\bigl\langle}(\pi\pi)^\OUT,M_K\bigl|{\cal H}_W\bigr| K^0\bigr\rangle.
\end{eqnarray}
This identity provides no new information beyond a check of our method.  The left- and right-hand sides of Eq.~\eqref{eq:check} are essentially identical except that the states appearing on the left, $|\alpha,E{\rangle}^\infty$, must be normalized to a delta function in the energy while the states on the right, $|(\pi\pi)^\OUT,E=M_K{\bigr\rangle}^\infty$ are normalized according to Eq.~\eqref{eq:pipi_norm}, a difference for which the extra factor of $\omega_q q/16\pi^2$ on the right-hand side compensates.   This factor can be recognized by comparing the normalization conventions in Eq.~\eqref{eq:pipi_norm} with conventions that are consistent with the generalized sum over $\alpha$ and energy used in Eq.~\eqref{eq:mixing_infty}:
\begin{equation}
\langle l'm', \vec P',E'|lm,\vec P, E\rangle = (2\pi)^3\delta^3(\vec P' - \vec P) \delta (E'-E)\delta_{l'l} \delta_{m'm},
\label{eq:E_norm}
\end{equation}
where for the generalized index $\alpha$ we have used the usual discrete angular momentum variables $lm$ and the center-of-mass momentum $\vec P$.

The equation of interest results after Eq.~\eqref{eq:check} is subtracted from Eq.~\eqref{eq:DM_2}, leaving an equation for the principal part alone:
\begin{eqnarray}
{\cal P}\int_{2m_\pi}^\infty dE \sum_\alpha 
\frac{\langle \overline{K^0}|H_W|\alpha,E{\rangle}^\infty 
         \prescript{\infty}{}{\langle} \alpha,E|{\cal H}_W| K^0\rangle}
        {M_K-E} 
&=& \sum_n \frac{\langle \overline{K^0}|H_W|n{\rangle}^\V 
         \prescript{\V}{}{\langle}n|H_W| K^0\rangle}
        {M_K-E_n} 
\label{eq:DM_3} \\ \nonumber
&& \hskip -3.0 in - \frac{\omega_q q}{16\pi^2}\cot\bigl(\phi(M_K)+\delta_0(M_K)\bigr)
\bigl\langle \overline{K^0}\bigl|{\cal H}_W\bigr|(\pi\pi)^\OUT,M_K{\bigr\rangle}^\infty 
                   \prescript{\infty}{}{\bigl\langle}(\pi\pi)^\OUT,M_K\bigl|{\cal H}_W\bigr| K^0\bigr\rangle \\ \label{eq:DM_4}
&&\hskip -3.5 in = \sum_n \frac{\langle \overline{K^0}|H_W|n{\rangle}^\V 
         \prescript{\V}{}{\langle}n|H_W| K^0\rangle}
        {M_K-E_n} 
\label{eq:DM_4}  - \cot\bigl(\phi(M_K)+\delta_0(M_K)\bigr)\left. \frac{d\bigl(\phi(E)+\delta_0(E)\bigr)}{dE}\right|_{E=M_K}
 \\ \nonumber
 && \hskip -1.5 in \cdot\bigl\langle \overline{K^0}\bigl|H_W\bigr|\pi\pi ,M_K{\bigr\rangle}^{\V'} 
                   \prescript{\V'}{}{\bigl\langle}\pi\pi,M_K\bigl|H_W\bigr| K^0\bigr\rangle.
\end{eqnarray}
Equation~\eqref{eq:DM_4} can be written in the more compact from:
\begin{eqnarray}
M_{\overline{0}0}^\infty = M_{\overline{0}0}^\V  - \cot\bigl(\phi(M_K)+\delta_0(M_K)\bigr)
                     \left. \frac{d\bigl(\phi(E)+\delta_0(E)\bigr)}{dE}\right|_{E=M_K}
 \label{eq:DM_5} \\ \nonumber
 && \hskip -1.5 in       \cdot\bigl\langle \overline{K^0}\bigl|H_W\bigr|\pi\pi,M_K{\bigr\rangle}^{\V'} 
                   \prescript{\V'}{}{\bigl\langle}\pi\pi,M_K\bigl|H_W\bigr| K^0\bigr\rangle.
\end{eqnarray}
Our primary result, Eq.~\eqref{eq:DM_3}, for the finite-volume correction to $\Delta M_K$ is given directly in terms of infinite-volume quantities.   However, to be useful in a finite-volume lattice calculation we need to express the correction in terms of finite-volume quantities that also can be computed using lattice methods.  This is done in Eqs.~\eqref{eq:DM_4} and \eqref{eq:DM_5}.  Nevertheless, we should recognize that the explicit, finite-volume $K\to\pi\pi$ matrix element, $\langle \overline{K^0}\bigl|H_W\bigr|\pi\pi,M_K{\bigr\rangle}^{\V'} $, which appears in Eqs.~\eqref{eq:DM_4} and \eqref{eq:DM_5} is energy conserving and must be evaluated using an appropriate volume $V'$, adjusted to ensure that $E_{\pi\pi}=M_K$, which may be different from the volume $V$ being used to compute $M_{\overline{0}0}^\V$.
  
It is reassuring to note that as the finite volume is adjusted so that an energy $E_n$ of a discrete, finite-volume state approaches $M_K$ the $1/(M_K-E_n)$ pole in the finite-volume quantity $M_{\overline{0}0}^\V$ will be canceled by a corresponding pole in the factor $\cot\bigl(\phi(M_K)+\delta_0(M_K)\bigr) d\bigl(\phi(E)+\delta_0(E)\bigr)/dE$.  In fact, we can examine the singular case where $E_n\to M_K$ in which the $1/(M_K-E_n)$ term in $M_{\overline{0}0}^\V$ is canceled.  Expanding the right-hand side of Eq.~\eqref{eq:DM_5} to zeroth order in $M_K-E_n$ for that case gives the formula:
\begin{eqnarray}
M_{\overline{0}0}^\infty &=& \left(M_{\overline{0}0}^\V\right)^\prime - \frac{d}{dE}\left\{ \bigl\langle \overline{K^0}\bigl|H_W\bigr|\pi\pi,E{\bigr\rangle}^\V 
                   \prescript\V{}{\bigl\langle}\pi\pi,E\bigl|H_W\bigr| K^0\bigr\rangle
\right\} \Bigr|_{E=M_K} 
\label{eq:DM_6}  \\
&&\hskip 0.2 in +\frac{1}{2} \frac{1}{\frac{d(\phi+\delta_0)}{dE}}
                           \frac{d^2(\phi+\delta_0)}{d E^2}
\bigl\langle \overline{K^0}\bigl|H_W\bigr|\pi\pi,E{\bigr\rangle}^\V 
                   \prescript\V{}{\bigl\langle}\pi\pi,E\bigl|H_W\bigr| K^0\bigr\rangle
\nonumber
\end{eqnarray}
where the prime superscript on $M_{\overline{0}0}^\V$ indicates that the state $|n\rangle$ with $E_n = M_K$ has been omitted from the sum in Eq.~\eqref{eq:mixing_FV}. 
The second term on the right-hand side of Eq.~\eqref{eq:DM_6} results from expanding the numerator in $M_{\overline{0}0}^V$ to first order in $M-E_n$ while the third term comes from a similar expansion of argument of the cotangent in Eq.~\eqref{eq:DM_5}.  This special case, where the volume is chosen so that $E_n=M_K$, was the first result that was obtained for the finite-volume corrections to $M_{\overline{0}0}$~\cite{Christ:2010zz,Christ:2012np} using a generalization of Lellouch and L\"ushcer's indirect approach of examining the resonant structure of $\pi\pi$ scattering that results when the effects of the second-order weak interactions and $K_L$ and $K_S$ intermediate states are studied.   While we have chosen to derive the result given in 
Eq.~\eqref{eq:DM_5} using the methods of KSS, this result can also be obtained using the techniques developed in Ref.~\cite{Lin:2001ek} and it was this approach 
which lead us to the more general formula given in Eq.~\eqref{eq:DM_5} and first presented in Ref.~\cite{Christ:2014qaa}.

It should be noted that Eq.~\eqref{eq:DM_5} relates the complete complex off-diagonal mixing matrix $M_{\overline{0}0}$ computed in finite and infinite volume: both its real and imaginary parts.  Thus, it can be used in both the calculation of $\Delta M_K$ and $\epsilon_K$.  This contrasts to the indirect approach to these relations taken in Refs.~\cite{Christ:2010zz} and \cite{Christ:2012np} where a very different strategy was needed for these two cases.

\section{Conclusion}
\label{sec:conclusion}

In this paper we have derived the finite-volume corrections necessary to determine
the $K_L -K_S$ mass difference and the long-distance contributions to $\epsilon_K$ from lattice simulations in an Euclidean finite volume.  Because of the principal part which appears in the infinite-volume expression for  $M_{\overline{0}0}$ and the potential singular behavior of the corresponding finite-volume quantity, quantitative control over these finite-volume effects is necessary before a finite-volume lattice QCD calculation of $M_{\overline{0}0}$ can be related to the infinite-volume $K_L-K_S$ mixing matrix.  In fact, the relation between the finite- and infinite-volume quantities is quite simple and the necessary finite-volume correction can also be evaluated if the finite-volume $K\to\pi\pi$ matrix elements can be computed.  Such matrix elements are necessarily a part of a calculation of $M_{\overline{0}0}$ since the two-pion state will typically also lead to unphysical terms which grow exponentially as the temporal region over which the two weak Hamiltonians are integrated grows; for example, see Eq.~(2) of Ref.~\cite{Bai:2014cva}.  Thus, $K\to\pi\pi$ matrix element must already be included in a calculation of $M_{\overline{0}0}$ to remove these unwanted terms.  However, the matrix element needed for this finite-volume correction is evaluated for energy-conserving kinematics with $E_{\pi\pi}=M_K$.  This may require additional calculation with a different volume chosen to achieve $E_{\pi\pi}=M_K$, a singular choice that is likely best to avoid for the actual calculation of $M_{\overline{0}0}$.  Fortunately, exploratory calculations~\cite{Bai:2014hta} suggest that the two-pion contribution to $M_{\overline{0}0}$ is quite small so even an approximate estimate of this on-shell decay amplitude may be adequate even for an accurate calculation of $M_{\overline{0}0}$.

As noted above the result for the finite-volume correction for $M_{\overline{0}0}$ is an important component of a first-principles calculation of the long-distance contribution to both $\Delta M_K$ and $\epsilon_K$.  (Here by long distance we refer to phenomena at or below the scale of the charm quark mass.)  The calculation of these quantities with systematic errors on the order of 10\% may be a practical five-year goal.  Exploratory calculations~\cite{Christ:2012se, Bai:2014cva,Bai:2014hta} which include all graphs suggest that statistical errors can be controlled at the few percent level, including those which arise from disconnected diagrams and that the approach to pions with physical mass also does not pose insurmountable difficulties.  The largest obstacle to a complete calculation of $M_{\overline{0}0}$ with controlled errors is the need to include both the heavy charm quark, which requires a small lattice spacing, and physical pions, which requires a large volume.  The RBC and UKQCD collaborations are presently generating an $80^2\times96\times192$ ensemble with an inverse lattice spacing $1/a\approx 3$ GeV and physical values for the light, strange and charm quark masses.  However, it is likely that a complete calculation of $M_{\overline{0}0}$ on this and ensembles with finer lattice spacing must wait for the next generation of high-performance computers.

After this work was completed we learned of an independent analysis~\cite{Briceno:2015csa} in which it has also been recognized that the KSS approach can be used to provide a third, alternative derivation of the Lellouch-L\"uscher relation between infinite- and finite-volume decay matrix elements, as has been shown in Section~\ref{sec:LL} above, supplementing those in Refs.~\cite{Lellouch:2000pv} and \cite{Lin:2001ek}. 

\section{Acknowledgement}
C.T.S., N.H.C.~and X.F.~would like to thank their colleagues in the RBC and UKQCD collaboration for helpful discussions.  N.H.C. and X.F. were supported in part by US DOE Grant No.DE-SC0011941 and C.T.S.~by UK STFC Grants ST/G000557/1 and ST/L000296/1. G.M. was supported in part by the ERC-2010 DaMESyFla Grant Agreement Number: 267985 and the MUIR (Italy) under a contract PRIN10.

\bibliography{references}

\begin{thebibliography}{21}%
\makeatletter
\providecommand \@ifxundefined [1]{%
 \@ifx{#1\undefined}
}%
\providecommand \@ifnum [1]{%
 \ifnum #1\expandafter \@firstoftwo
 \else \expandafter \@secondoftwo
 \fi
}%
\providecommand \@ifx [1]{%
 \ifx #1\expandafter \@firstoftwo
 \else \expandafter \@secondoftwo
 \fi
}%
\providecommand \natexlab [1]{#1}%
\providecommand \enquote  [1]{``#1''}%
\providecommand \bibnamefont  [1]{#1}%
\providecommand \bibfnamefont [1]{#1}%
\providecommand \citenamefont [1]{#1}%
\providecommand \href@noop [0]{\@secondoftwo}%
\providecommand \href [0]{\begingroup \@sanitize@url \@href}%
\providecommand \@href[1]{\@@startlink{#1}\@@href}%
\providecommand \@@href[1]{\endgroup#1\@@endlink}%
\providecommand \@sanitize@url [0]{\catcode `\\12\catcode `\$12\catcode
  `\&12\catcode `\#12\catcode `\^12\catcode `\_12\catcode `\%12\relax}%
\providecommand \@@startlink[1]{}%
\providecommand \@@endlink[0]{}%
\providecommand \url  [0]{\begingroup\@sanitize@url \@url }%
\providecommand \@url [1]{\endgroup\@href {#1}{\urlprefix }}%
\providecommand \urlprefix  [0]{URL }%
\providecommand \Eprint [0]{\href }%
\providecommand \doibase [0]{http://dx.doi.org/}%
\providecommand \selectlanguage [0]{\@gobble}%
\providecommand \bibinfo  [0]{\@secondoftwo}%
\providecommand \bibfield  [0]{\@secondoftwo}%
\providecommand \translation [1]{[#1]}%
\providecommand \BibitemOpen [0]{}%
\providecommand \bibitemStop [0]{}%
\providecommand \bibitemNoStop [0]{.\EOS\space}%
\providecommand \EOS [0]{\spacefactor3000\relax}%
\providecommand \BibitemShut  [1]{\csname bibitem#1\endcsname}%
\let\auto@bib@innerbib\@empty
%</preamble>
\bibitem [{\citenamefont {Olive}\ \emph {et~al.}(2014)\citenamefont {Olive}
  \emph {et~al.}}]{Agashe:2014kda}%
  \BibitemOpen
  \bibfield  {author} {\bibinfo {author} {\bibfnamefont {K.~A.}\ \bibnamefont
  {Olive}} \emph {et~al.} (\bibinfo {collaboration} {Particle Data Group}),\
  }\href@noop {} {\bibfield  {journal} {\bibinfo  {journal} {Chin.Phys.}\
  }\textbf {\bibinfo {volume} {C38}},\ \bibinfo {pages} {090001} (\bibinfo
  {year} {2014})}\BibitemShut {NoStop}%
%%CITATION = CHPHD,C38,090001;%%
\bibitem [{\citenamefont {Brod}\ and\ \citenamefont
  {Gorbahn}(2012)}]{Brod:2011ty}%
  \BibitemOpen
  \bibfield  {author} {\bibinfo {author} {\bibfnamefont {J.}~\bibnamefont
  {Brod}}\ and\ \bibinfo {author} {\bibfnamefont {M.}~\bibnamefont {Gorbahn}},\
  }\href {\doibase 10.1103/PhysRevLett.108.121801} {\bibfield  {journal}
  {\bibinfo  {journal} {Phys.Rev.Lett.}\ }\textbf {\bibinfo {volume} {108}},\
  \bibinfo {pages} {121801} (\bibinfo {year} {2012})},\ \Eprint
  {http://arxiv.org/abs/1108.2036} {arXiv:1108.2036 [hep-ph]} \BibitemShut
  {NoStop}%
%%CITATION = ARXIV:1108.2036;%%
\bibitem [{\citenamefont {Christ}(2010)}]{Christ:2010zz}%
  \BibitemOpen
  \bibfield  {author} {\bibinfo {author} {\bibfnamefont {N.~H.}\ \bibnamefont
  {Christ}} (\bibinfo {collaboration} {RBC and UKQCD Collaborations}),\
  }\href@noop {} {\bibfield  {journal} {\bibinfo  {journal} {PoS}\ }\textbf
  {\bibinfo {volume} {LATTICE2010}},\ \bibinfo {pages} {300} (\bibinfo {year}
  {2010})}\BibitemShut {NoStop}%
%%CITATION = POSCI,LATTICE2010,300;%%
\bibitem [{\citenamefont {Yu}(2011)}]{Yu:2011gk}%
  \BibitemOpen
  \bibfield  {author} {\bibinfo {author} {\bibfnamefont {J.}~\bibnamefont
  {Yu}},\ }\href@noop {} {\bibfield  {journal} {\bibinfo  {journal} {PoS}\
  }\textbf {\bibinfo {volume} {LATTICE2011}},\ \bibinfo {pages} {297} (\bibinfo
  {year} {2011})},\ \Eprint {http://arxiv.org/abs/1111.6953} {arXiv:1111.6953
  [hep-lat]} \BibitemShut {NoStop}%
%%CITATION = ARXIV:1111.6953;%%
\bibitem [{\citenamefont {Christ}(2011)}]{Christ:2012np}%
  \BibitemOpen
  \bibfield  {author} {\bibinfo {author} {\bibfnamefont {N.~H.}\ \bibnamefont
  {Christ}},\ }\href@noop {} {\bibfield  {journal} {\bibinfo  {journal} {PoS}\
  }\textbf {\bibinfo {volume} {LATTICE2011}},\ \bibinfo {pages} {277} (\bibinfo
  {year} {2011})},\ \Eprint {http://arxiv.org/abs/1201.2065} {arXiv:1201.2065
  [hep-lat]} \BibitemShut {NoStop}%
%%CITATION = ARXIV:1201.2065;%%
\bibitem [{\citenamefont {Yu}(2012)}]{Yu:2012nx}%
  \BibitemOpen
  \bibfield  {author} {\bibinfo {author} {\bibfnamefont {J.}~\bibnamefont
  {Yu}},\ }\href@noop {} {\bibfield  {journal} {\bibinfo  {journal} {PoS}\
  }\textbf {\bibinfo {volume} {LATTICE2012}},\ \bibinfo {pages} {129} (\bibinfo
  {year} {2012})},\ \Eprint {http://arxiv.org/abs/1212.0234} {arXiv:1212.0234
  [hep-lat]} \BibitemShut {NoStop}%
%%CITATION = ARXIV:1212.0234;%%
\bibitem [{\citenamefont {Christ}\ \emph {et~al.}(2013)\citenamefont {Christ},
  \citenamefont {Izubuchi}, \citenamefont {Sachrajda}, \citenamefont {Soni},\
  and\ \citenamefont {Yu}}]{Christ:2012se}%
  \BibitemOpen
  \bibfield  {author} {\bibinfo {author} {\bibfnamefont {N.}~\bibnamefont
  {Christ}}, \bibinfo {author} {\bibfnamefont {T.}~\bibnamefont {Izubuchi}},
  \bibinfo {author} {\bibfnamefont {C.}~\bibnamefont {Sachrajda}}, \bibinfo
  {author} {\bibfnamefont {A.}~\bibnamefont {Soni}}, \ and\ \bibinfo {author}
  {\bibfnamefont {J.}~\bibnamefont {Yu}} (\bibinfo {collaboration} {RBC and
  UKQCD Collaborations}),\ }\href {\doibase 10.1103/PhysRevD.88.014508}
  {\bibfield  {journal} {\bibinfo  {journal} {Phys.Rev.}\ }\textbf {\bibinfo
  {volume} {D88}},\ \bibinfo {pages} {014508} (\bibinfo {year} {2013})},\
  \Eprint {http://arxiv.org/abs/1212.5931} {arXiv:1212.5931 [hep-lat]}
  \BibitemShut {NoStop}%
%%CITATION = ARXIV:1212.5931;%%
\bibitem [{\citenamefont {Yu}(2013)}]{Yu:2013qfa}%
  \BibitemOpen
  \bibfield  {author} {\bibinfo {author} {\bibfnamefont {J.}~\bibnamefont
  {Yu}},\ }\href@noop {} {\bibfield  {journal} {\bibinfo  {journal} {PoS}\
  }\textbf {\bibinfo {volume} {LATTICE2013}},\ \bibinfo {pages} {398} (\bibinfo
  {year} {2013})},\ \Eprint {http://arxiv.org/abs/1312.0306} {arXiv:1312.0306
  [hep-lat]} \BibitemShut {NoStop}%
%%CITATION = ARXIV:1312.0306;%%
\bibitem [{\citenamefont {Bai}\ \emph {et~al.}(2014)\citenamefont {Bai},
  \citenamefont {Christ}, \citenamefont {Izubuchi}, \citenamefont {Sachrajda},
  \citenamefont {Soni} \emph {et~al.}}]{Bai:2014cva}%
  \BibitemOpen
  \bibfield  {author} {\bibinfo {author} {\bibfnamefont {Z.}~\bibnamefont
  {Bai}}, \bibinfo {author} {\bibfnamefont {N.}~\bibnamefont {Christ}},
  \bibinfo {author} {\bibfnamefont {T.}~\bibnamefont {Izubuchi}}, \bibinfo
  {author} {\bibfnamefont {C.}~\bibnamefont {Sachrajda}}, \bibinfo {author}
  {\bibfnamefont {A.}~\bibnamefont {Soni}},  \emph {et~al.},\ }\href {\doibase
  10.1103/PhysRevLett.113.112003} {\bibfield  {journal} {\bibinfo  {journal}
  {Phys.Rev.Lett.}\ }\textbf {\bibinfo {volume} {113}},\ \bibinfo {pages}
  {112003} (\bibinfo {year} {2014})},\ \Eprint {http://arxiv.org/abs/1406.0916}
  {arXiv:1406.0916 [hep-lat]} \BibitemShut {NoStop}%
%%CITATION = ARXIV:1406.0916;%%
\bibitem [{\citenamefont {Luscher}(1986)}]{Luscher:1986pf}%
  \BibitemOpen
  \bibfield  {author} {\bibinfo {author} {\bibfnamefont {M.}~\bibnamefont
  {Luscher}},\ }\href@noop {} {\bibfield  {journal} {\bibinfo  {journal}
  {Commun. Math. Phys.}\ }\textbf {\bibinfo {volume} {105}},\ \bibinfo {pages}
  {153} (\bibinfo {year} {1986})}\BibitemShut {NoStop}%
%%CITATION = CMPHA,105,153;%%
\bibitem [{\citenamefont {Luscher}(1991)}]{Luscher:1990ux}%
  \BibitemOpen
  \bibfield  {author} {\bibinfo {author} {\bibfnamefont {M.}~\bibnamefont
  {Luscher}},\ }\href@noop {} {\bibfield  {journal} {\bibinfo  {journal} {Nucl.
  Phys.}\ }\textbf {\bibinfo {volume} {B354}},\ \bibinfo {pages} {531}
  (\bibinfo {year} {1991})}\BibitemShut {NoStop}%
%%CITATION = NUPHA,B354,531;%%
\bibitem [{\citenamefont {Hansen}\ and\ \citenamefont
  {Sharpe}(2012)}]{Hansen:2012tf}%
  \BibitemOpen
  \bibfield  {author} {\bibinfo {author} {\bibfnamefont {M.~T.}\ \bibnamefont
  {Hansen}}\ and\ \bibinfo {author} {\bibfnamefont {S.~R.}\ \bibnamefont
  {Sharpe}},\ }\href {\doibase 10.1103/PhysRevD.86.016007} {\bibfield
  {journal} {\bibinfo  {journal} {Phys.Rev.}\ }\textbf {\bibinfo {volume}
  {D86}},\ \bibinfo {pages} {016007} (\bibinfo {year} {2012})},\ \Eprint
  {http://arxiv.org/abs/1204.0826} {arXiv:1204.0826 [hep-lat]} \BibitemShut
  {NoStop}%
%%CITATION = ARXIV:1204.0826;%%
\bibitem [{\citenamefont {Lellouch}\ and\ \citenamefont
  {Luscher}(2001)}]{Lellouch:2000pv}%
  \BibitemOpen
  \bibfield  {author} {\bibinfo {author} {\bibfnamefont {L.}~\bibnamefont
  {Lellouch}}\ and\ \bibinfo {author} {\bibfnamefont {M.}~\bibnamefont
  {Luscher}},\ }\href@noop {} {\bibfield  {journal} {\bibinfo  {journal}
  {Commun. Math. Phys.}\ }\textbf {\bibinfo {volume} {219}},\ \bibinfo {pages}
  {31} (\bibinfo {year} {2001})},\ \Eprint
  {http://arxiv.org/abs/hep-lat/0003023} {hep-lat/0003023} \BibitemShut
  {NoStop}%
%%CITATION = HEP-LAT 0003023;%%
\bibitem [{\citenamefont {Lin}\ \emph {et~al.}(2001)\citenamefont {Lin},
  \citenamefont {Martinelli}, \citenamefont {Sachrajda},\ and\ \citenamefont
  {Testa}}]{Lin:2001ek}%
  \BibitemOpen
  \bibfield  {author} {\bibinfo {author} {\bibfnamefont {C.~J.~D.}\
  \bibnamefont {Lin}}, \bibinfo {author} {\bibfnamefont {G.}~\bibnamefont
  {Martinelli}}, \bibinfo {author} {\bibfnamefont {C.~T.}\ \bibnamefont
  {Sachrajda}}, \ and\ \bibinfo {author} {\bibfnamefont {M.}~\bibnamefont
  {Testa}},\ }\href@noop {} {\bibfield  {journal} {\bibinfo  {journal} {Nucl.
  Phys.}\ }\textbf {\bibinfo {volume} {B619}},\ \bibinfo {pages} {467}
  (\bibinfo {year} {2001})},\ \Eprint {http://arxiv.org/abs/hep-lat/0104006}
  {hep-lat/0104006} \BibitemShut {NoStop}%
%%CITATION = HEP-LAT 0104006;%%
\bibitem [{\citenamefont {Rummukainen}\ and\ \citenamefont
  {Gottlieb}(1995)}]{Rummukainen:1995vs}%
  \BibitemOpen
  \bibfield  {author} {\bibinfo {author} {\bibfnamefont {K.}~\bibnamefont
  {Rummukainen}}\ and\ \bibinfo {author} {\bibfnamefont {S.~A.}\ \bibnamefont
  {Gottlieb}},\ }\href@noop {} {\bibfield  {journal} {\bibinfo  {journal}
  {Nucl. Phys.}\ }\textbf {\bibinfo {volume} {B450}},\ \bibinfo {pages} {397}
  (\bibinfo {year} {1995})},\ \Eprint {http://arxiv.org/abs/hep-lat/9503028}
  {hep-lat/9503028} \BibitemShut {NoStop}%
%%CITATION = HEP-LAT 9503028;%%
\bibitem [{\citenamefont {Christ}\ \emph {et~al.}(2005)\citenamefont {Christ},
  \citenamefont {Kim},\ and\ \citenamefont {Yamazaki}}]{Christ:2005gi}%
  \BibitemOpen
  \bibfield  {author} {\bibinfo {author} {\bibfnamefont {N.~H.}\ \bibnamefont
  {Christ}}, \bibinfo {author} {\bibfnamefont {C.}~\bibnamefont {Kim}}, \ and\
  \bibinfo {author} {\bibfnamefont {T.}~\bibnamefont {Yamazaki}},\ }\href
  {\doibase 10.1103/PhysRevD.72.114506} {\bibfield  {journal} {\bibinfo
  {journal} {Phys. Rev.}\ }\textbf {\bibinfo {volume} {D72}},\ \bibinfo {pages}
  {114506} (\bibinfo {year} {2005})},\ \Eprint
  {http://arxiv.org/abs/hep-lat/0507009} {arXiv:hep-lat/0507009} \BibitemShut
  {NoStop}%
%%CITATION = HEP-LAT/0507009;%%
\bibitem [{\citenamefont {Kim}\ \emph {et~al.}(2005)\citenamefont {Kim},
  \citenamefont {Sachrajda},\ and\ \citenamefont {Sharpe}}]{Kim:2005gf}%
  \BibitemOpen
  \bibfield  {author} {\bibinfo {author} {\bibfnamefont {C.~h.}\ \bibnamefont
  {Kim}}, \bibinfo {author} {\bibfnamefont {C.~T.}\ \bibnamefont {Sachrajda}},
  \ and\ \bibinfo {author} {\bibfnamefont {S.~R.}\ \bibnamefont {Sharpe}},\
  }\href@noop {} {\bibfield  {journal} {\bibinfo  {journal} {Nucl. Phys.}\
  }\textbf {\bibinfo {volume} {B727}},\ \bibinfo {pages} {218} (\bibinfo {year}
  {2005})},\ \Eprint {http://arxiv.org/abs/hep-lat/0507006} {hep-lat/0507006}
  \BibitemShut {NoStop}%
%%CITATION = HEP-LAT 0507006;%%
\bibitem [{\citenamefont {Donoghue}\ \emph {et~al.}(1992)\citenamefont
  {Donoghue}, \citenamefont {Golowich},\ and\ \citenamefont
  {Holstein}}]{Donoghue:1992dd}%
  \BibitemOpen
  \bibfield  {author} {\bibinfo {author} {\bibfnamefont {J.~F.}\ \bibnamefont
  {Donoghue}}, \bibinfo {author} {\bibfnamefont {E.}~\bibnamefont {Golowich}},
  \ and\ \bibinfo {author} {\bibfnamefont {B.~R.}\ \bibnamefont {Holstein}},\
  }\href@noop {} {\bibfield  {journal} {\bibinfo  {journal} {Camb. Monogr.
  Part. Phys. Nucl. Phys. Cosmol.}\ }\textbf {\bibinfo {volume} {2}},\ \bibinfo
  {pages} {1} (\bibinfo {year} {1992})}\BibitemShut {NoStop}%
%%CITATION = CMPCE,2,1;%%
\bibitem [{\citenamefont {Christ}\ \emph {et~al.}(2014)\citenamefont {Christ},
  \citenamefont {Martinelli},\ and\ \citenamefont
  {Sachrajda}}]{Christ:2014qaa}%
  \BibitemOpen
  \bibfield  {author} {\bibinfo {author} {\bibfnamefont {N.~H.}\ \bibnamefont
  {Christ}}, \bibinfo {author} {\bibfnamefont {G.}~\bibnamefont {Martinelli}},
  \ and\ \bibinfo {author} {\bibfnamefont {C.~T.}\ \bibnamefont {Sachrajda}},\
  }\href@noop {} {\bibfield  {journal} {\bibinfo  {journal} {PoS}\ }\textbf
  {\bibinfo {volume} {LATTICE2013}},\ \bibinfo {pages} {399} (\bibinfo {year}
  {2014})},\ \Eprint {http://arxiv.org/abs/1401.1362} {arXiv:1401.1362
  [hep-lat]} \BibitemShut {NoStop}%
%%CITATION = ARXIV:1401.1362;%%
\bibitem [{\citenamefont {Bai}(2014)}]{Bai:2014hta}%
  \BibitemOpen
  \bibfield  {author} {\bibinfo {author} {\bibfnamefont {Z.}~\bibnamefont
  {Bai}},\ }\href@noop {} {\  (\bibinfo {year} {2014})},\ \Eprint
  {http://arxiv.org/abs/1411.3210} {arXiv:1411.3210 [hep-lat]} \BibitemShut
  {NoStop}%
%%CITATION = ARXIV:1411.3210;%%
\bibitem [{\citenamefont {Briceno}\ and\ \citenamefont
  {Hansen}(2015)}]{Briceno:2015csa}%
  \BibitemOpen
  \bibfield  {author} {\bibinfo {author} {\bibfnamefont {R.~A.}\ \bibnamefont
  {Briceno}}\ and\ \bibinfo {author} {\bibfnamefont {M.~T.}\ \bibnamefont
  {Hansen}},\ }\href@noop {} {\  (\bibinfo {year} {2015})},\ \Eprint
  {http://arxiv.org/abs/1502.04314} {arXiv:1502.04314 [hep-lat]} \BibitemShut
  {NoStop}%
%%CITATION = ARXIV:1502.04314;%%
\end{thebibliography}%

\end{document}